# A Novel Approach Towards Identification of Alcohol and Drug Induced People

Joyjit Chatterjee, Anita Thakur, Vajja Mukesh
Amity School of Engineering and Technology
Amity University, U.P, Noida-201303, India
joyjit.c.in@ieee.org, athakur@amity.edu, mukesh.vajja@gmail.com

*Abstract*—**The paper proposes a novel approach towards identification of alcohol and drug induced people, through the use of a wearable bracelet.As alcohol and drug induced human people are in an unconsciousstate of mind, they need external help from the surroundings.With proposed Bracelet system we can identify the alcohol and drug indused people and warning trigger message is sent to their care takers. There is a definite relationship between an individual's Blood Alcohol Content (BAC) and Pulse Rate to identify the alcohol or drug consumed person .This relationship of pulse rate with BAC is sensed by piezoelectric sensor and warning system is developed as a Bracelet device . The viability of the Bracelet is verified by Simulating a Database of 199 People's BAC and Pulse Rate Features and classification is done among the Alcohol Induced and Normal People. For classification,Ensemble Boosted Tree Algorithm is used which is having 81.9% accuracy in decision.**

## I. INTRODUCTION

Every day, one can hear of multiple incidents in the media, where people, especially women, are subjected to drug and alcohol laced drinks at parties or other such occasions. These people unfortunately become victims of heinous crimes such are rapes, murders and sexual harassment. By developing a simple, yet efficient device in the form of a bracelet, it can become much easier to keep track of such incidents and report the crime to the emergency contacts, including the Police, to take appropriate measures to save the person from the life-threatening situation.

The literature survey into existing work in this area reveals that presently, most of the available smart devices for identification of alcohol and drug induced are expensive and use complex hardware. In a recent work by Ben Lovejoy,wearable infrared sensors are used for the purpose of identification of pulse rate, which is thereby used as the key predictor for detecting alcohol or drug induced situation [1]. An ongoing prototype being developed by a US Company, Milo Sensors Ltd. useswearable sensors for analyzing and interpreting the Breath Alcohol Content information for recognition of an individual[2]. The proposed bracelet is different as it makes use of a embedded piezoelectric sensor. Moreover, unlike the complex hardware required in [2], the proposed bracelet makes use of a simple low-cost microcontroller, which makes critical decisions based on the Arterial Pulse Rate of the individual. Instead of using Alcohol Levels directly in human breath, it uses the Arterial Pulse Rate as the vital parameter.

A reputed company, Scram Systems, proposed a fairly new work in this area, they make use of wearable sensors for analyzing and interpreting the ankle-based alcohol information for recognition of an individual involved in a crime[3]. However, we propose a bracelet which doesn't use complex ankle-based sensors. Our proposed bracelet sends information about the person's location based on the GPS Coordinates to the nearest police station and the emergency contacts, whereas, the work in [3] simply predicts the recognition at user-end, without any external communication capabilities.

To establish the viability of the proposed bracelet in recognizing the alcohol and drug induced people, a database of 199 people's BAC and Pulse Rate is used as predictors, and then, based on the values of these predictors, classification is done among the Alcohol Induced and the Normal people. By the application of MATLAB Classification Learner Toolbox, the Boosted Tree Algorithm gives the best possible result, with an accuracy of 81.9% in decision making.

Organization of paper as follows: Section II explains the proposed hardware design.Section III describes the methodology used.Section IV discusses the database adopted. Section V discusses the results with the analysis of the proposed features. Section VI concludes the paperand futureaspect of proposed work.

## II. PROPOSED HARDWARE DESIGN

The Emergency Trigger Bracelet consists of a Piezoelectric Sensor, which senses the pulse rate of the individual through the bracelet worn on the wrist. Fig 1 shows the proposed system block diagram. Piezoelectric Transducer is convert the pulse rate into an electrical signal then it is transmitted to a noise cancellation circuit. The noise-cancelled signal is transmitted to the Microcontroller. Where Microcontroller is facilitated with provision of a Sound Sensor Module and Bluetooth Module. Whenever the human pulse is out of range of the normal range of pulse, resulting from alcohol or drug overdose, or, the help sound is uttered by the user, the activation signal is sent from the microcontroller to the Bluetooth module. The In-Built GPS Module of the User's Mobile provides the appropriate location coordinates of the user, which is finally sent as an SMS with GPS Location Coordinates to the Police & Emergency Contacts, thus, intimating them of an emergency situation.

## III METHODOLOGY

The Proposed Bracelet makes use of Piezoelectric Sensor for sensing the Arterial Pulse Rate of an individual, who is wearing the bracelet, and this Pulse Rate is used as a Measure of Alcohol Saturation Levels in the individual. We propose the Bracelet which is in-vitro (used outside human body).Thus, the proposed Bracelet in non invasive and painless. The bracelet which uses a low-cost microcontroller, which performs all the decision making based on the Arterial Pulse Levels. Our Proposed Project makes use of a Noise Cancellation Electronic Circuit, which makes it more precise and reliable. The Bracelet has an embedded Sound Sensor, which keeps on continuously monitoring for "Help" sound utterance by the individual. This makes the Proposed Bracelet more reliable, and provides an additional safety measure for intimating emergency, just in case the Piezoelectric Pulse Sensor fails in sensing abnormal Pulse Rates. The Proposed Bracelet makes use of a piezoelectric sensor to sense the Arterial Pulse Rate of a person, taken through the person's wrist. The Emergency Trigger Bracelet which doesn't have expensive components like Ultrasonic Sensors, Cameras etc.is well suited for day-to-day use by users and realization in the Industry[4]. Process of working of proposed system explained in flow chart in figure 2.

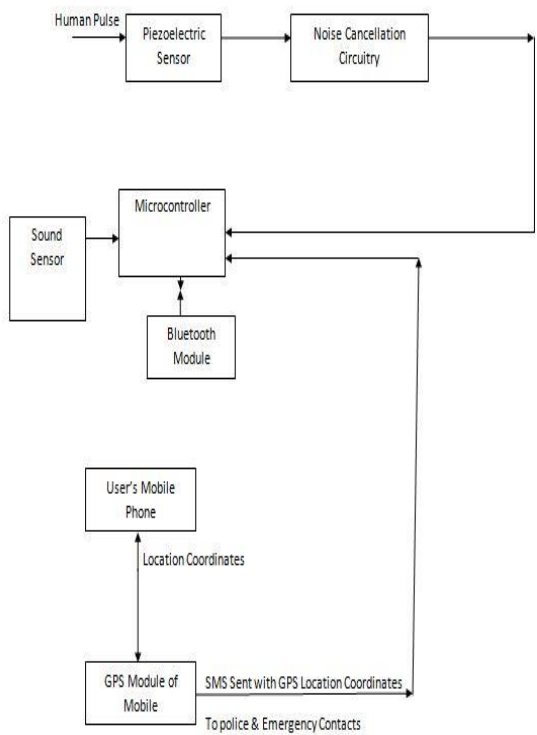

Figure 1. Proposed system Hardware Design

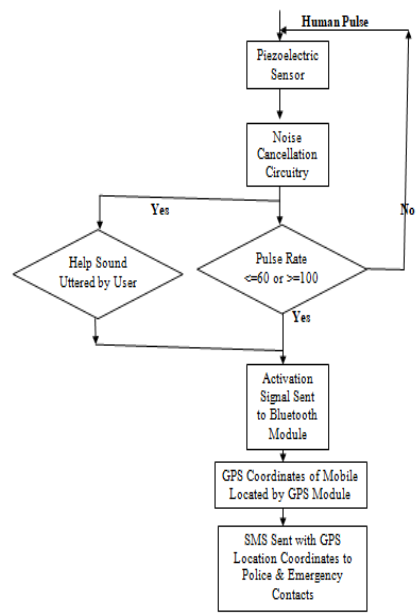

Figure 2. Flow Chart of the Proposed system

The Process is elaborated in detail below:-
1. Initially, the Human Pulse is Continually Sensed by the Piezoelectric Sensor.
2. The Piezoelectric Sensor converts the Human Pulse Signal into an electrical Signal, which is then transmitted to the Noise Cancellation Circuitry.
3. The Noise Cancelled Signalis sent to the Microcontroller, and based on the decision control of threshold value of human pulse due to alcohol or drug overdose, or due to help sound being uttered by the user, the activation signal is sent from the microcontroller to the Bluetooth module. If the situation is negative feedback case, then, the closed loop negative feedback is sent back and Human Pulse again measured continually.
4. Then, the GPS Location of the User's Mobile is located by its built-in GPS Module.
5. Finally, an SMS is sent with GPS Location Coordinates to the Police and the Individuals Emergency Contacts, to intimate them of a possible emergency situation.

For the Purpose of testing the viability of the proposed algorithm, the BAC and Pulse Rate are used as 2 predictors, further concatenated into a 199x3 Matrix with the third column as the target (0 for Normal and 1 for Drug Induced), which is fed to the MATLAB's Classification Learner Toolbox. Five Fold Cross-Validation is used and the Alcohol Induced and Normal People are classified using various classifiers [5,6].The accuracy of multiple classifiers are checked in parallel to identify the best possible classifier for the purpose of this paper. Figure 3 highlights the process for testing the accuracy of the proposed algorithm.

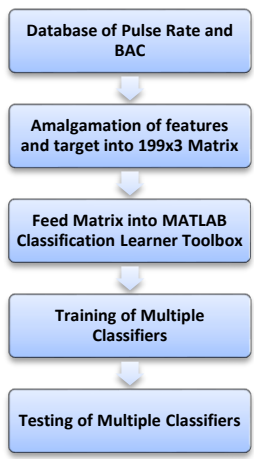

Figure 3. Process for Testing Accuracy of Proposed Algorithm

IV. DATABASE

In the paper, a database of 199 people's BAC and Pulse Rate is collected from the PhysionNet, the research resource for complex physiologic signals [7]. An individual's BAC and Pulse Rate are used as two predictors for the purpose of classification of alcohol induced and normal person in this paper. Figure 4 describes the 199x3 Matrix as a part of the Database used for this paper.

| | 1 BAC | 2 PulseRate | 3 Target |
|---|---|---|---|
| 1 | 0 | 0 | 0 |
| 2 | 0 | 0.0500 | 0 |
| 3 | 0 | 0.2500 | 0 |
| 4 | 0 | 0.3500 | 0 |
| 5 | 0 | 0.4500 | 1 |
| 6 | 0 | 0.5000 | 1 |
| 7 | 0 | 0.5500 | 0 |
| 8 | 0 | 0.6000 | 1 |
| 9 | 0 | 0.7500 | 1 |
| 10 | 0 | 0.8000 | 1 |
| 11 | 0 | 0.8500 | 0 |
| 12 | 0 | 0.9000 | 1 |
| 13 | 0 | 1.1000 | 1 |
| 14 | 0 | 1.1500 | 1 |
| 15 | 0 | 1.3000 | 1 |
| 16 | 0 | 1.3500 | 1 |
| 17 | 0 | 1.5000 | 0 |
| 18 | 0 | 1.5500 | 0 |
| 19 | 0 | 1.7000 | 0 |
| 20 | 0 | 1.9000 | 1 |
| 21 | 0 | 1.9500 | 1 |

Figure 4. 199x3 Matrix as a part of Database used

V. RESULTS

When a person is consumed with alcohol or drug pulse rate and other physiological activities are changed [8,9]. That showed in proposed paper work with classifier.The Figure 5 describes the Scatter Diagram, which clearly shows a large distinction between Alcohol Induced and Normal Person's BAC and Pulse Rate.

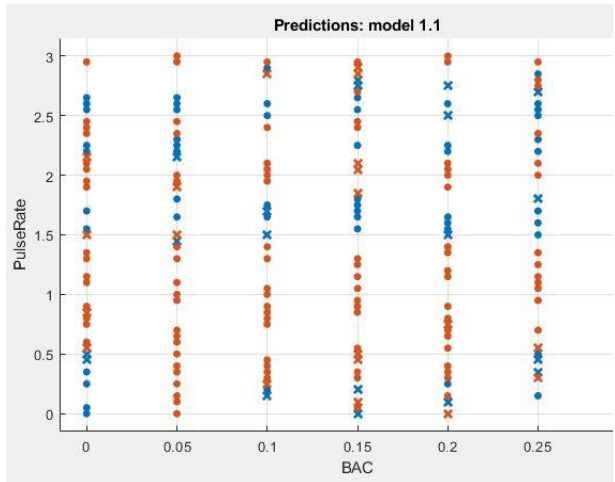

Figure 5. Scatter Diagram for BAC and Pulse Rate

It can be clearly seen that the BAC and Pulse rate are integral physiological signals, which when used as features can help identify alcohol and drug induced person [10,11]. The abnormal pulse rate reaches to a standard deviation of 3 and the abnormal BAC reaches a value of 0.25 in case a person is alcohol or drug induced, as is evident from the Scatter Diagram in Figure 5.

Figure 6 demonstrates the Confusion Matrix for the Database used in this paper.

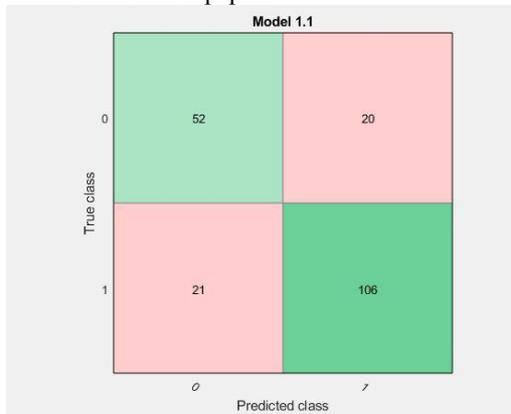

Figure 6. 199x3 Matrix as a part of Database used

Figure 7 shows the Confusion Matrix between the Predicted Class along with the True Positive Rate and False Negative Rate obtained after Classification.

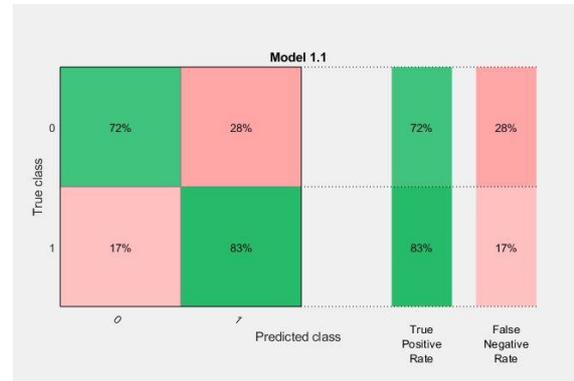

Figure 7. Confusion Matrix with TP and FP Rates

Figure 8 clearly highlights the Confusion Matrix between the True Class (Indicated by 1) and the Positive Predictive Value and False Discovery Rate.

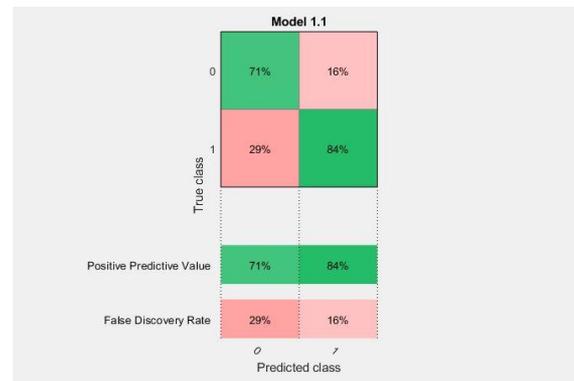

Figure 8. Confusion Matrix between True Class and PPV and FDR

Figure 9 shows the Parallel Coordinates Plot obtained using the MATLAB Classification Learner Toolbox. The two predictors, namely BAC and Pulse Rate are clearly demonstrated along the visual demonstration of the number of correct and incorrect predictions during classification.

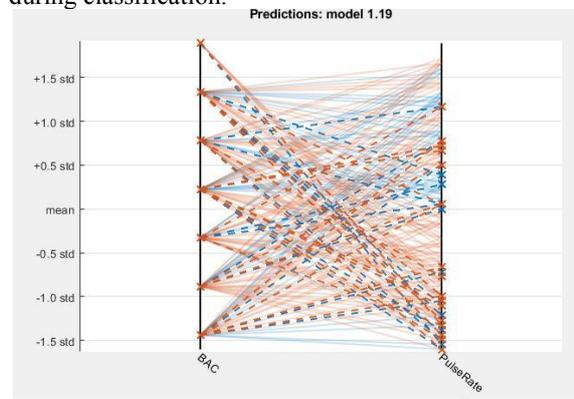

Figure 9. Parallel Coordinates Plot between BAC and Pulse Rate Predictors

Table 1 shows the MATLAB Classification Learner Toolbox's accuracy for various classifiers which were tested in parallel using fivefold cross-validation.

Table 1: Performance Evaluation of Various Classifiers used

| Classifier Type | Accuracy Obtained (%) |
|---|---|
| Fine Tree | 79.4 |
| Medium Tree | 78.4 |
| Coarse Tree | 63.8 |
| Linear Discriminant | 60.3 |
| Quadratic Discriminant | 62.3 |
| Logistic Regression | 60.3 |
| Linear SVM | 63.8 |
| Quadratic SVM | 63.8 |
| Cubic SVM | 59.3 |
| Fine Gaussian SVM | 63.3 |
| Medium Gaussian SVM | 59.3 |
| Coarse Gaussian SVM | 63.8 |
| Fine KNN | 70.9 |
| Medium KNN | 65.8 |
| Coarse KNN | 63.8 |
| Cosine KNN | 57.3 |
| Cubic KNN | 65.3 |
| Weighted KNN | 70.4 |
| **Boosted Trees** | **81.9** |
| Bagged Trees | 78.9 |
| Subspace Discriminant | 63.8 |
| Subspace KNN | 64.3 |
| RUSBoosted Trees | 78.4 |

The highlighted value of 81.9% accuracy is the optimal value obtained using the Ensemble Boosted Trees Classifier as shown in Figure 10.

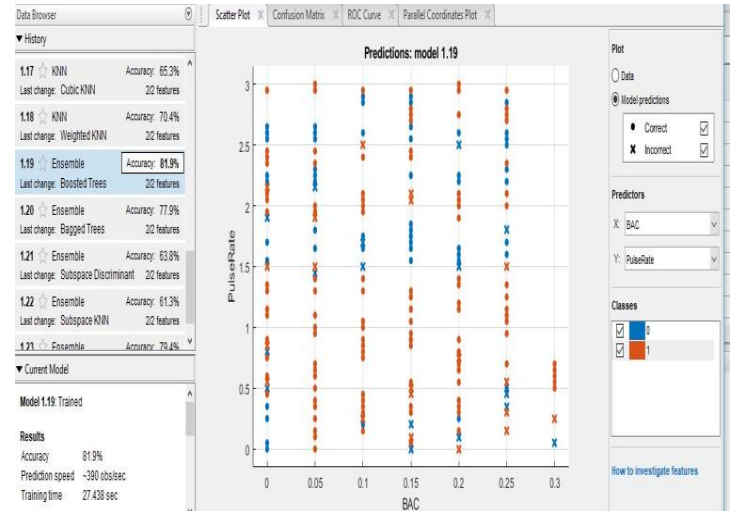

Figure 10. Accuracy obtained using MATLAB Classification Learner Toolbox

## VI. CONCLUSION

The proposed system is of utmost application to the public in general, women in particular and the Police at large. As the system uses the pulse rate measurement through surface electrode for the complete analysis and prediction of whether or not a person is alcohol or drug induced, it is completely reliable and safe. Furthermore, the device resembles a simple bracelet, which can be worn by men or women alike and one can't easily identify the purpose of the device without opening it, thus preventing the criminals from guessing the same. Also, as the road accidents are increasing sharply nowadays, the bracelet will help to curb such incidents by insisting the user and informing the police about such situations. The miniaturized SIM card and GPS modules placed within the bracelet can provide a wealth of information and evidence for assisting the Police and the person's family in case a mishap occurs.

Moreover, the viability of the proposed algorithm in the bracelet, making use of BAC and Pulse Rate as predictors for classification between normal and Alcohol induced gives a good accuracy of 81.9% using the Ensemble Boosted Tree Classifier. The proposed bracelet can be implemented according to the hardware design mentioned, and act as a smart device, which is cost efficient and help identify the alcohol induced people. It will also serve to reduce crime towards women and other people in the near future.